\documentclass[prc,superscriptaddress,twocolumn,showpacs,preprintnumbers]{revtex4}

\usepackage[dvips]{graphicx}
\usepackage{amsmath}
\usepackage{amssymb}
\usepackage{ulem}
\usepackage{times}
\usepackage{mathrsfs}
\usepackage{multirow}
\usepackage{bm}

\newcommand{\beq}{\begin{equation}}
\newcommand{\eeq}{\end{equation}}
\newcommand{\bea}{\begin{eqnarray}}
\newcommand{\eea}{\end{eqnarray}}

\def\beq{\begin{equation}}
\def\eeq{\end{equation}}
\def\bea{\begin{eqnarray}}
\def\eea{\end{eqnarray}}
\def\fun#1#2{\lower3.6pt\vbox{\baselineskip0pt\lineskip.9pt
  \ialign{$\mathsurround=0pt#1\hfil##\hfil$\crcr#2\crcr\sim\crcr}}}

\begin{document}

\title{Anti-halo effects on reaction cross sections 
for $^{14,15,16}$C isotopes
}

\author{Takuma Matsumoto}
\affiliation{Department of Physics, Kyushu University, Fukuoka 812-8581, Japan}

\author{Masanobu Yahiro}
\affiliation{Department of Physics, Kyushu University, Fukuoka 812-8581, Japan}

\date{\today}

\begin{abstract} 
We study anti-halo effects on reaction cross sections 
$\sigma_{\rm R}$ for $^{14,15,16}$C scattering 
from a $^{12}$C target at 83~MeV/nucleon, 
using the $g$-matrix double-folding model. 
$^{15}$C is described by the $^{14}$C~+~$n$ two-body model 
that reproduces the measured large s-wave spectroscopic factor, i.e., 
the shell inversion that the 1s$_{1/2}$ orbital is lower 
than the 0d$_{5/2}$ orbital in energy. 
$^{16}$C is described by the $^{14}$C~+~$n$~+~$n$ 
three-body model with the phenomenological three-body force (3BF) that 
explains the measured small s-wave spectroscopic factor. 
The 3BF allows the single-particle energies 
of the $^{14}$C + $n$ subsystem to depend 
on the position $r$ of the second neutron from the center of mass 
of the subsystem. 
The 1s$_{1/2}$ orbital is lower than the 0d$_{5/2}$ orbital for large $r$, 
but the shell inversion is restored for small $r$. 
Anti-halo effects due to the ``partial shell inversion'' 
make $\sigma_{\rm R}$ for $^{16}$C smaller than that for $^{15}$C. 
We also investigate projectile breakup effects on the mass-number dependence 
of $\sigma_{\rm R}$ 
with the continuum discretized coupled-channels method.  
\end{abstract}


\maketitle
\section{Introduction}
Unstable nuclei have exotic properties such as 
halo formation~\cite{Tanihata85,Ozawa01} 
and shell evolution; see Ref.~\cite{Sorlin:2014pba} for the recent review on 
shell evolution. 
Elucidation of these properties is an important subject in nuclear physics. 
Reaction cross section $\sigma_{\rm R}$ is a powerful experimental tool 
for determining matter radii of nuclei and hence searching for halo nuclei. 
In addition, theoretical analyses for $\sigma_{\rm R}$ are easier
compared with other reactions. In fact, $\sigma_{\rm R}$ was measured
recently for the scattering of Ne and Mg isotopes from a $^{12}$C  
at 240~MeV/nucleon~\cite{Takechi12,Takechi-Mg}, and the double-folding
model (DFM) based on the Melbourne $g$-matrix \cite{Amos} was successful
in reproducing the data with no free parameter
\cite{Minomo:2011sj,Minomo:2011bb,Sumi:2012fr,Watanabe:2014zea}. The
analyses suggest that $^{31}$Ne and $^{37}$Mg are halo nuclei with large
deformation.

Pairing correlations are known to be important for even nuclei.  
The correlation plays an important role particularly in weakly bound nuclei, 
since they are bound only with it. 
In the mean-field picture based on the Hartree-Fock Bogoliubov (HFB)
method~\cite{Bennaceur00}, the correlation makes quasi-particle
energies deeper and then reduces the root-mean-square (RMS) radius of the
matter density. This mechanism becomes significant for 
unstable nuclei where the separation energy is smaller 
than the gap energy.  This suggests that the pairing correlation 
suppresses halo formation for even-even unstable nuclei. 
This is called the pairing anti-halo effect.

The pairing anti-halo effect is an interesting mechanism, but there is no 
clear evidence for the effect. Hagino and Sagawa suggested that the 
odd-even staggering in $\sigma_{R}$ is possible  
evidence for the effect~\cite{Hagino:2011ji,Hagino:2011aa,Hagino:2012qu}, 
using the HFB method for $^{30,31,32}$Ne~+~$^{12}$C scattering at
240~MeV/nucleon~\cite{Takechi12}  and the few-body models for
$^{14,15,16}$C~+~$^{12}$C scattering 
at 83~MeV/nucleon~\cite{Fang04}. 
They introduced the parameter \cite{Hagino:2012qu} 
\bea
\gamma_3=\sigma_{\rm R}(A+1)-\frac{\sigma_{\rm R}(A+2) +\sigma_{\rm R}(A)}{2}, 
\label{staggering-parameter}
\eea
where the mass number $A$ of projectile is even 
in Eq. \eqref{staggering-parameter}.  
The parameter $\gamma_3$ describes the deviation of 
$\sigma_{\rm R}(A+1)$ for an odd nucleus 
from the mean value $(\sigma_{\rm R}(A+2) +\sigma_{\rm R}(A))/2$ 
for even nuclei on both sides. 
Sasabe {\it et al.} extended their idea and defined 
the dimensionless odd-even deviation parameter~\cite{Sasabe:2013dwa} 
\bea
\Gamma_{\rm R}&=&\frac{\gamma_3}{[\sigma_{\rm R}(A+2)
-\sigma_{\rm R}(A)]/2} ,     
\label{OES-para-2}
\eea
where 
$\Gamma_{\rm R} >1$ when 
$\sigma_{\rm R}(A+1) > \sigma_{\rm R}(A+2)$. 
The parameter evaluated from measured $\sigma_{\rm R}$ 
has a large value of $\Gamma_{\rm R}^{\rm exp}=2.0 \pm 0.8$ 
for $^{14,15,16}$C~+~$^{12}$C scattering at 83 MeV/nucleon~\cite{Fang04}. 
The fact that $\Gamma_{\rm R}^{\rm exp} > 1$ shows that 
anti-halo effects play an important role in $\sigma_{\rm R}$ 
for $^{16}${\rm C}. Sasabe {\it et al.} \cite{Sasabe:2013dwa} 
analyzed the strong odd-even deviation  
with the continuum-discretized coupled-channels method (CDCC) 
\cite{Kam86,Aus87,Yah12} in order to 
take account of projectile breakup effects in addition to 
pairing (di-neutron) correlations. 
Here we identify di-neutron correlations with pairing ones.
In the analysis, $^{15}$C was described by the $^{14}$C + $n$ two-body 
orthogonality condition model (OCM) 
and $^{16}$C was by the $^{14}$C + $n$ + $n$ three-body OCM 
with the phenomenological three-body force (3BF) 
that reproduces the total binding energy. The theoretical calculations well 
reproduce $\sigma_{\rm R}$ for $^{14,15}$C, but the calculated odd-even 
deviation parameter is $\Gamma_{\rm R}=0.77$ and significantly undershoots 
$\Gamma_{\rm R}^{\rm exp}=2.0 \pm 0.8$, although pairing  
correlations are taken into account in the three-body model. 
This implies that there exist other anti-halo effects besides 
the pairing anti-halo effect.

$^{15}$C is a halo nucleus with small one-neutron separation energy, 
$S_n =1.218$~MeV. The s-wave spectroscopic factor $\lambda_s$ in the 
ground state of $^{15}$C is found 
to be $\lambda_s(^{15}{\rm C}) = 0.97 \pm 0.08$ 
from Coulomb breakup 
measurements \cite{Pramanik2003}. 
This means that the 1s$_{1/2}$ orbital is lower than the 0d$_{5/2}$ orbital 
in energy, i.e., the shell inversion takes place. 
The s-wave spectroscopic factor in the ground state of $^{16}$C, meanwhile, 
is determined to be $\lambda_s(^{16}{\rm C})=0.35 \pm 0.2$ 
from the measurements on longitudinal momentum 
distributions of $^{15}$C fragments from $^{16}$C breakup 
\cite{Yamaguchi2003}. 
The suppression of $\lambda_s$ 
from $\lambda_s(^{15}{\rm C})= 0.97 \pm 0.08$ to 
$\lambda_s(^{16}{\rm C})=0.35 \pm 0.2$ 
is a key 
to clarifying the reason why $\Gamma_{\rm R}^{\rm exp}$ is so large.

In this paper, we reanalyze $^{14,15,16}$C~+~$^{12}$C scattering at
83~MeV/nucleon by using  the Melbourne $g$-matrix DFM 
and study the mechanism underlying the strong odd-even  
deviation of $\Gamma_{\rm R}^{\rm exp}=2.0 \pm 0.8$. 
We introduce a surface-type 3BF 
to the $^{14}$C + $n$ + $n$ three-body model 
to describe the suppression of $\lambda_s$ 
from $\lambda_s(^{15}{\rm C})$ to $\lambda_s(^{16}{\rm C})$. 
The 3BF allows the single-particle energies 
of the $^{14}$C + $n$ subsystem to depend 
on the distance $r$ between the second neutron and the center of mass 
of the $^{14}$C + $n$ 
subsystem. The 1s$_{1/2}$ orbital is lower than the 0d$_{5/2}$
orbital for large $r$, but the shell inversion is restored for small $r$. 
This is referred to as ``partial shell inversion'' in this paper. 
We show that the partial shell inversion enhances $\Gamma_{\rm R}$ 
largely. 
We also investigate projectile breakup effects on $\Gamma_{\rm R}$ 
with CDCC.

We briefly explain the DFM for nucleus--nucleus scattering 
and the few-body models for $^{15,16}$C in Sec. \ref{Model setting} 
and show the results of model calculations in Sec. \ref{Results}. 
Section \ref{Summary} is devoted to summary.

\section{Model setting}
\label{Model setting}
In the present $g$-matrix DFM, 
the optical potential $U$ for nucleus--nucleus scattering 
is obtained by folding the Melbourne $g$-matrix 
with projectile and target densities; see Ref.~\cite{Sumi:2012fr} 
for the detail. 
The reaction cross section is obtained by solving the one-body Schr\"odinger 
with $U$ for the elastic $S$-matrix elements. 
For a $^{12}$C target, the matter density is assumed to be identical with 
the proton density deduced from the electron scattering~\cite{C12-density}, 
since the proton RMS radius deviates from 
the neutron one only by less than 1\% in HFB calculations 
with the Gogny-D1S interaction~\cite{GognyD1S}. 
For $^{14}$C, the matter density is determined by HFB calculations, where 
the center-of-mass correction is made in the standard 
manner~\cite{Sumi:2012fr}. 
The matter radius of $^{14}$C in the HFB calculation 
is ${\bar r}$($^{14}$C)$=2.51$ fm 
that is consistent with the measured charge radius 2.50 fm~\cite{Sc82e}.

As for $^{15}$C, we use the $^{14}$C + $n$ two-body 
OCM in which 
the Pauli-forbidden states are excluded in the modelspace~\cite{Saito69}. 
The Hamiltonian is 
\begin{eqnarray}
 h_2&=&T_{\rho} + V_{n{\rm c}} ,
\end{eqnarray}
where $T_{\rho}$ is the kinetic-energy operator with respect to 
the relative coordinate $\rho$ between $n$ and the core nucleus ($^{14}$C). 
The interaction $V_{n{\rm c}}$ between $n$ and $^{14}$C is taken from
Ref.~\cite{Hagino:2011ji}. 
In this model, the shell inversion takes place, that is, 
the 1s$_{1/2}$ orbital is lower than the 0d$_{5/2}$ orbital. 
As a consequence of this property, this model well reproduces 
properties of the ground and 1st-excited states of $^{15}$C such as 
$\lambda_s (^{15}{\rm C})= 0.97 \pm 0.08$.  
The matter radius of $^{15}$C predicted by this model 
is ${\bar r}$($^{15}$C)$=2.87$ fm that is much larger than 
${\bar r}$($^{14}$C)$=2.51$ fm. 
If the shell inversion does not take place, the calculated radius becomes 
${\bar r}$($^{15}$C)$=2.65$ fm that is estimated by assuming the ground
state with 0d$_{5/2}$ and $S_n=1.218$ MeV. 
The shell inversion is thus inevitable for $^{15}$C to have a halo structure.

As for $^{16}$C, we use the $^{14}$C~+~$n$~+~$n$ three-body OCM. 
The Hamiltonian is 
\bea
 h_3&=&T_{\rho_1}+T_{r_1}+V, 
\label{eq:h3}
\eea
which consists of the kinetic-energy operators $T_{\rho_1}$ and $T_{r_1}$ 
with respect to two Jacobi coordinates and the interaction $V$ defined by 
\bea
 V&=&V_{n_{1}n_{2}}+V_{n_{1}{\rm c}}+V_{n_{2}{\rm c}}+V_{3}, 
\eea
where $V_{n_{1}n_{2}}$ is the two-nucleon force acting 
between two valence neutrons, $n_{1}$ and $n_{2}$, 
and $V_{n_{1}{\rm c}}$ ($V_{n_{2}{\rm c}}$) 
is the interaction between $n_{1}$ ($n_{2}$) and $^{14}$C. 
We use the Bonn-A two-nucleon force~\cite{Mac89} as $V_{n_{1}n_{2}}$ and 
the nucleon--$^{14}$C interaction of  Ref.~\cite{Hagino:2011ji} 
as $V_{n_{1}{\rm c}}$ and $V_{n_{2}{\rm c}}$.  
The interaction $V_{3}$ is the 3BF acting among $n_{1}$, $n_{2}$, and 
$^{14}$C. We consider two types of 3BFs. One is a {\it volume-type 3BF} of 
\bea
V_{3}^{(v)}=\sum_{c=1,2}V_{0}^{(v)} e^{-(\rho_c/\rho_0)^2} e^{-(r_c/r_0)^2} , 
\eea
and the other is a {\it surface-type 3BF} of 
\bea
V_{3}^{(s)}=\sum_{c=1,2}V_{0}^{(s)} 
\rho_c^2 e^{-(\rho_c/\rho_0)^2} e^{-(r_c/r_0)^2},
\eea
where $\rho_1$ ($\rho_2$) is the coordinate of $n_{1}$ ($n_{2}$) from $^{14}$C 
and $r_1$ ($r_2$) represents the coordinate of $n_{2}$ ($n_{1}$) from 
the center of mass of the $n_1$ ($n_2$)~+~$^{14}$C subsystem. 
Assuming $\rho_0=0.76 \times 14^{1/3}$ fm 
and $r_0=2.54 \times 14^{1/3}$ fm, 
we determined $V_{0}^{(v)}$ and $V_{0}^{(s)}$ from the measured two-neutron
separation energy $S_{2n}({\rm exp})=5.47$ MeV~\cite{Til93}; 
the resultant values are 
$V_{0}^{(v)}=-23.45$ MeV and $V_{0}^{(s)}=-6.18$ MeV/fm$^{2}$. 
Note that $T_{\rho_1}+T_{r_1}=T_{\rho_2}+T_{r_2}$ 
in Eq. \eqref{eq:h3}.

For later convenience, we define the following four models by changing $V$: 
(I) $V=V_{n_{1}{\rm c}}+V_{n_{2}{\rm c}}$,
(II) $V=V_{n_{1}n_{2}}+V_{n_{1}{\rm c}}+V_{n_{2}{\rm c}}$,
(III) $V=V_{n_{1}n_{2}}+V_{n_{1}{\rm c}}+V_{n_{2}{\rm c}}+V_{3}^{(v)}$, 
and 
(IV) $V=V_{n_{1}n_{2}}+V_{n_{1}{\rm c}}+V_{n_{2}{\rm c}}+V_{3}^{(s)}$. 
Using the four models, we calculate the probabilities 
$P_{\rm s}$ and $P_{\rm d}$ of 1s$_{1/2}$ and 0d$_{5/2}$ components in 
the ground state of $^{16}$C, the matter radius ${\bar r}$($^{16}$C), 
and $S_{2n}$. 
The probabilities are obtained by taking the overlap between 
the ground state of $^{16}$C and 
the 1s$_{1/2}$ and 0d$_{5/2}$ states of $^{15}$C. 
The difference between models II and III (IV) shows 
effects of the volume-type 
(surface-type) 3BF, whereas the difference between models 
I and  II describes effects of pairing correlations.

In actual calculations, we used 
the Gaussian expansion method~\cite{Hiy03} in which $h_2$ and $h_3$ are
diagonalized in a space spanned by Gaussian basis functions with geometric
progression range parameters.

\section{Results}
\label{Results}

\subsection{Ground-state properties of $^{16}$C}

Table~\ref{Table-probabilities} shows a comparison 
of four models for $P_{\rm s}$, $P_{\rm d}$, the ratio 
$\delta={\bar r}(^{16}{\rm C})/{\bar r}(^{15}{\rm C})$, and $S_{2n}$. 
In model I, $P_{\rm s}$ is almost
100\%, so that $^{16}$C has a halo structure of 
$\delta=1.09$. In model II where $V_{nn}$ is added to model I, 
$P_{\rm s}$ decreases to 79\% whereas $P_{\rm d}$ increases to 16\%, 
so that $\delta$ is reduced to 1.04. 
Thus pairing correlations surely 
reduce $\delta$, but the reduction is only 5\%. 
The value of $S_{2n}$ for model II is still smaller than 
the experimental value $S_{2n}({\rm exp})=5.47$ MeV. This indicates that 
the introduction of phenomenological 3BF is necessary. 
In model III where $V_{3}^{(v)}$ is added to 
model II, $P_{\rm s}$ and $P_{\rm d}$ 
are close to the results in model II. 
The reduction of $\delta$ due to $V_{3}^{(v)}$ is 
6\% and comparable with that due to pairing correlation. 
In model IV where $V_{3}^{(s)}$ is added to 
model II, however, $P_{\rm s}$ is largely reduced to 15\% that is 
consistent with the measured value $\lambda_s(^{16}{\rm C})=0.35 \pm 0.2$, 
so that $\delta$ is reduced from 1.04 to 0.95. 
The reduction of $\delta$ due to $V_{3}^{(s)}$ is thus 9\% and twice as much as that due to pairing correlation.

\begin{table}[htbp]
\caption{Comparison of four models 
for the probabilities $P_{\rm s}$, $P_{\rm d}$ 
of 1s$_{1/2}$ and 0d$_{5/2}$ components in the ground state of $^{16}$C, 
the ratio $\delta={\bar r}(^{16}{\rm C})/{\bar r}(^{15}{\rm C})$ and 
the two-neutron separation energy $S_{2n}$, 
and the odd-even deviation parameter $\Gamma_{\rm rds}$.
}
\begin{tabular}{c|ccccc}
\hline \hline
Models&$P_{\rm s}$ [\%]&$P_{\rm d}$ [\%]&
$\delta$&$S_{2n}$ [MeV]&$\Gamma_{\rm rds}$\\
\hline
 I&99&0&1.09&2.40&0.11\\
 II&79&16&1.04&3.22&0.52\\
 III&72
     &14&0.98&5.47&1.50\\
 IV&15
     &74&0.95&5.47&2.48\\
\hline \hline
\end{tabular}
\label{Table-probabilities}
\end{table}

\begin{figure}[htbp]
\includegraphics[width=0.35\textwidth,clip]{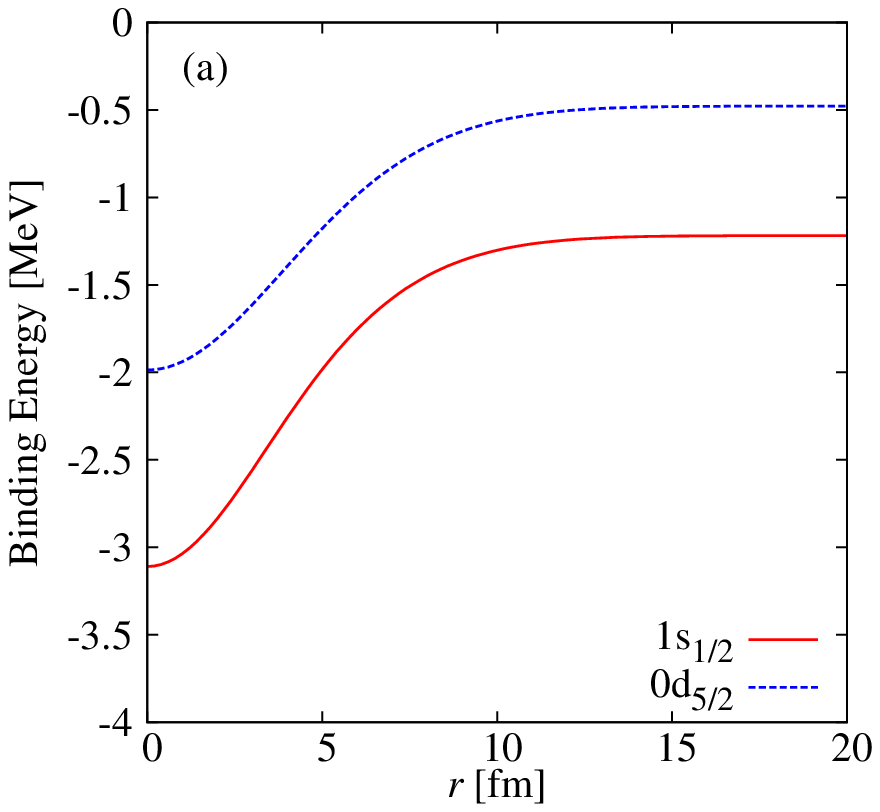}
\includegraphics[width=0.35\textwidth,clip]{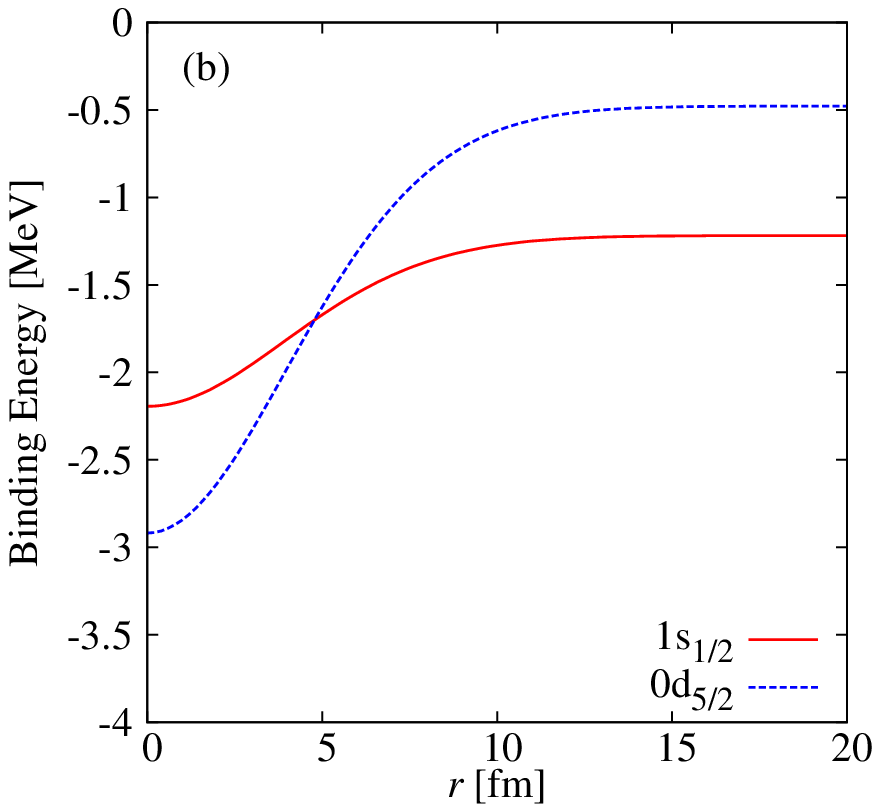}
\caption{(Color online) 
$r$ dependence of single particle energies in the $^{14}$C~+~$n$ subsystem 
of $^{16}$C. Panels (a) and (b) correspond to the results of 
the volume- and surface-type 3BFs, respectively. 
The solid (dashed) line corresponds to the 1s$_{1/2}$ (0d$_{5/2}$) orbital. 
}
\label{single-particle-energy}
\end{figure}

We make the following analysis to study the mechanism underlying 
the large reduction of $P_{\rm s}$ due to $V_{3}^{(s)}$. 
We start with the approximate Hamiltonian $T_{\rho_1}+V_{n_{1}\rm c}(\rho_1)$ 
instead of $h_3$. For the approximate Hamiltonian, one can clearly define 
the single-particle energies in the $^{14}$C~+~$n$ subsystem of $^{16}$C. 
The effects of 3BF on the single-particle energies can be estimated 
by adding a half of 3BF to the approximate Hamiltonian:  
\begin{eqnarray}
 h_3^\prime(r_1)&=&T_{\rho_1}+V_{n_{1}\rm c}(\rho_1)+V_{3}(\rho_1,r_{1}) ,
\label{eq:h3'}
\end{eqnarray}
where $V_{3}(\rho_1,r_{1})$ is the $c=1$ part of $V_3$ and 
the OCM is taken . 
The single-particle energies of $h_3^\prime(r_1)$ are obtained as a function 
of $r_1$. 
The single-particle energies are plotted as a function of $r$ for 
1s$_{1/2}$ and 0d$_{5/2}$ orbitals in Fig.~\ref{single-particle-energy}, 
where we have used $r$ as the shorthand notation of $r_1$. 
In panel (a) for the volume-type 3BF, the 1s$_{1/2}$ orbital is lower than 
the 0d$_{5/2}$ orbital for any $r$. The shell inversion thus takes place 
for any $r$. 
In panel (b) for the surface-type 3BF, 
meanwhile, the 1s$_{1/2}$ orbital is lower than the 0d$_{5/2}$ orbital 
at large $r$, but the shell inversion is restored at $r < 5$ fm.  
This partial shell inversion is an origin of the large reduction of 
$P_{\rm s}$ due to $V_{3}^{(s)}$.

Now we consider the sum ${\bar R}(A) = {\bar r}(A) + {\bar r}(A_{\rm T})$ 
of projectile and target RMS radii ${\bar r}(A)$ and ${\bar r}(A_{\rm T})$, 
and introduce the dimensionless odd-even deviation parameter 
$\Gamma_{\rm rds}$ for ${\bar R}(A)$ as 
\bea
\Gamma_{\rm rds} = 
\frac{{\bar R}^2(A+1) - [{\bar R}^2(A) + {\bar R}^2(A+2)]/2} 
{[{\bar R}^2(A+2) - {\bar R}^2(A)]/2} ,
\eea
where the mass number $A_{\rm T}$ of target is 12 in the present
case. The parameter $\Gamma_{\rm rds}$ has the same property as
$\Gamma_{\rm R}$; namely, 
$\Gamma_{\rm rds} > 1$ when 
${\bar R}(A+1) > {\bar R}(A+2)$.

The parameter $\Gamma_{\rm rds}$ is also tabulated 
in Table~\ref{Table-probabilities} for four models. 
In model I, $\Gamma_{\rm rds}$ is much smaller than 1. 
In model II where $V_{n_1n_2}$ is added to model I, 
$\Gamma_{\rm rds}$ becomes slightly large but still smaller than 1. 
Comparing models III and IV with model II, one can see that 
the enhancement of $\Gamma_{\rm rds}$ due to $V_{3}^{(v)}$ is 
much larger than that due to $V_{n_{1}n_{2}}$ and 
the enhancement of $\Gamma_{\rm rds}$ due to $V_{3}^{(s)}$ is even larger 
that that due to $V_{3}^{(v)}$. 
The partial shell inversion is thus important as an origin of 
the odd-even deviation in ${\bar R}(A)$, and $\Gamma_{\rm rds}$ is a good 
quantity to detect the partial shell inversion.

\subsection{Reaction cross sections}
The parameter $\Gamma_{\rm R}$ is identical with $\Gamma_{\rm rds}$, 
if the following two conditions are satisfied. 
The first condition is that projectile breakup is negligible. 
If it is significant, the effects on $\sigma_{\rm R}$ are larger 
for $^{15}$C projectile than for $^{16}$C projectile, since $^{15}$C has much 
smaller $S_n$ than $^{16}$C. This enhances $\Gamma_{\rm R}$ from 
$\Gamma_{\rm rds}$, but the following second condition is more significant 
in the present case. For simplicity, let us assume 
that projectile breakup is negligible.  
When the absolute value of the elastic $S$-matrix element is 0 
for orbital angular momenta $L$ corresponding to 
the nuclear interior and 1 for $L$ to the nuclear exterior, 
it is satisfied that $\sigma_{\rm R}(A) = \pi {\bar R}^2(A)$. 
This situation is called the black-sphere scattering (BSS). 
Hence $\Gamma_{\rm R}$ agrees with $\Gamma_{\rm rds}$, when 
the BSS is realized and projectile breakup is negligible. 
The present scattering largely differs from the BSS \cite{Sasabe:2013dwa}, 
so that $\Gamma_{\rm R}$ is much reduced from $\Gamma_{\rm rds}$ 
by the non-BSS effect, as shown below.  

\begin{figure}[htbp]
\includegraphics[width=0.35\textwidth,clip]{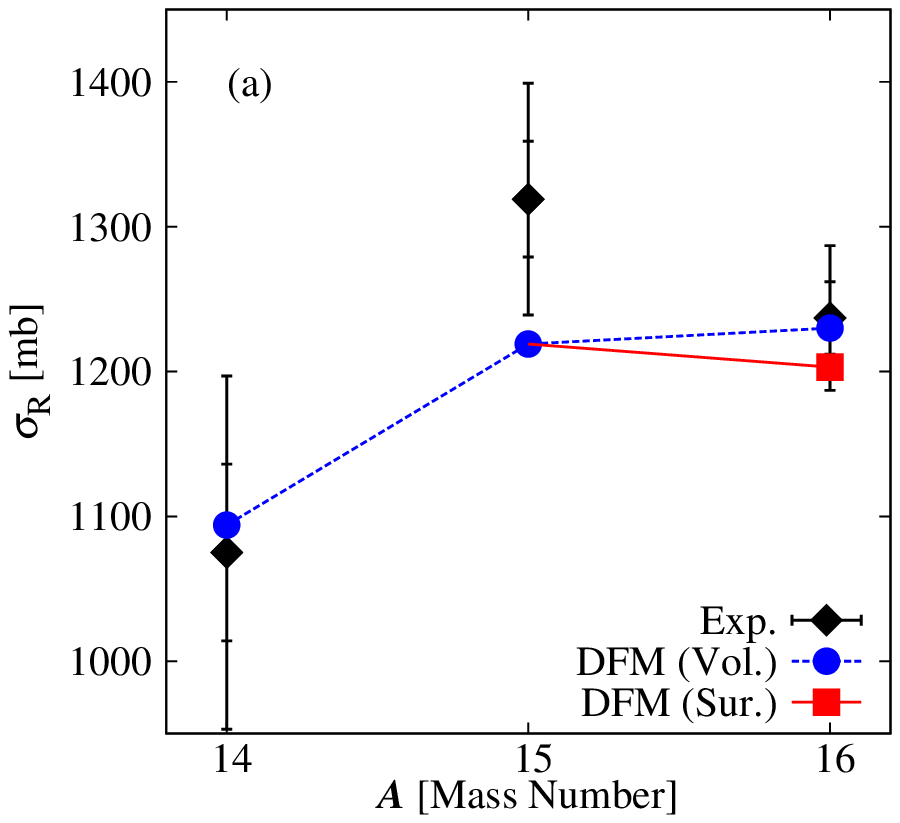}
\includegraphics[width=0.35\textwidth,clip]{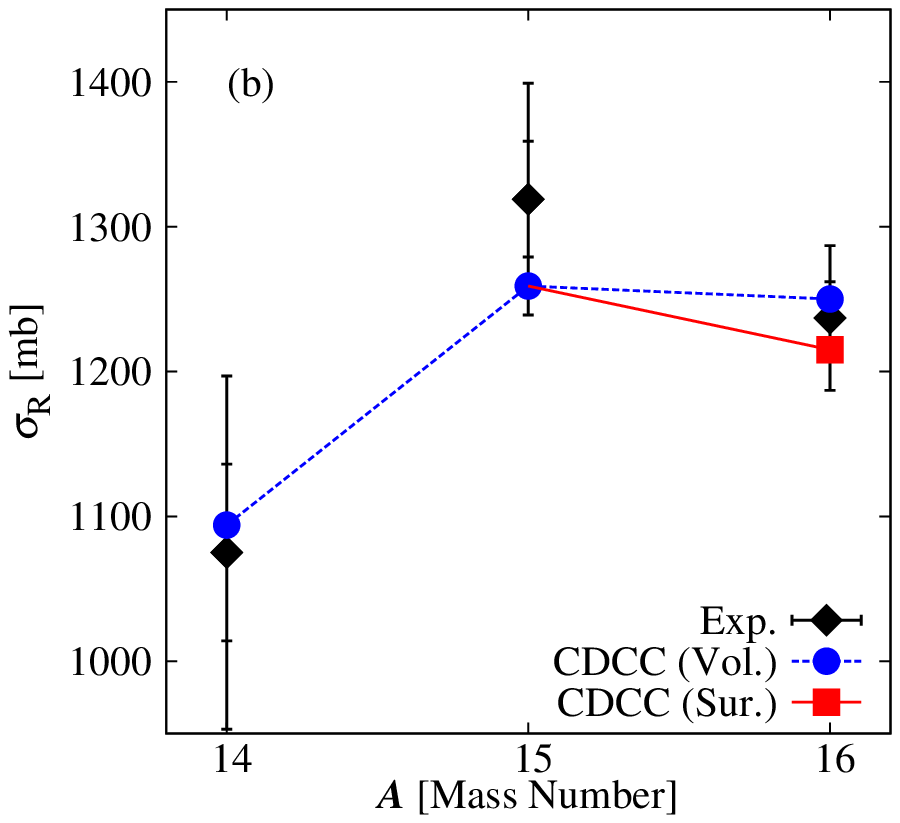}
\caption{(Color online)
Reaction cross sections $\sigma_{\rm R}$ for $^{14,15,16}$C~+~$^{12}$C 
scattering at 83 MeV/nucleon. Panels (a) and (b) represent the results of
 DFM and CDCC calculations, respectively. 
For $^{14,15}$C, the circles show the results of theoretical calculations. 
For $^{16}$C, the circle and square correspond 
to the results of the volume-type and surface-type 3BFs, respectively. 
The  experimental data are taken from  Ref.~\cite{Fang04} 
and are plotted  with 2-$\sigma$ error (95.4\% certainty).  
}
\label{RCS}
\end{figure}

It is not easy to estimate projectile breakup effects when 
the coupling potentials among elastic and breakup channels are calculated 
within the framework of the $g$-matrix double folding, 
since it requires time-consuming calculations.  
We then do CDCC calculations by assuming 
the $^{14}$C~+~$n$~+~$^{12}$C three-body model for $^{15}$C~+~$^{12}$C
scattering  and the $^{14}$C~+~$n$~+~$n$~+~$^{12}$C four-body model for
$^{16}$C~+~$^{12}$C scattering. The optical potentials $U_{x}$ between
$x(=n,^{14}$C) and a $^{12}$C target are constructed by folding the
Melbourne $g$-matrix with densities of $x$ and $^{12}$C. 
The optical potential $U_{n}$ is slightly reduced so that the single-channel 
calculation with no projectile breakup 
can yield the same $\sigma_{\rm R}$ as the DFM. 
The detail of CDCC calculations is the same as in Ref.~\cite{Sasabe:2013dwa}. 
Coulomb breakup is neglected, since it is small.  Convergence 
of CDCC solutions with respect to increasing 
the modelspace is confirmed for $\sigma_{\rm R}$.

Figure~\ref{RCS} shows $\sigma_{\rm R}$ for
$^{14,15,16}$C~+~$^{12}$C scattering  
at 83 MeV/nucleon. 
The parameter $\Gamma_{\rm R}$ is tabulated in Table 
\ref{Table-summary-staggering} for models III and IV.  
In panel (a) of Fig.~\ref{RCS}, the results of the DFM are 
plotted as a function of $A$. 
For $^{14,15}$C, the theoretical results are shown by circles.  
For $^{16}$C, the theoretical results are plotted by a circle 
for the volume-type 3BF and by a square for the surface-type 3BF. 
The parameter $\Gamma_{\rm R}$ is 1.24 and larger than 1 
for the surface-type 3BF, whereas $\Gamma_{\rm R}=0.79$ 
for the volume-type 3BF. The reaction cross sections thus have strong 
odd-even deviation for the surface-type 3BF. 
The reduction from $\Gamma_{\rm rds}=2.48$ 
to $\Gamma_{\rm R}=1.24$ for the surface-type 3BF 
is due to the non-BBS effect. 
In panel (b), projectile breakup corrections are added to the results of 
panel (a). The results well reproduce the measured $\sigma_{\rm R}$ for 
$^{14,15,16}$C~\cite{Fang04}; 
here the data are plotted 
with 2-$\sigma$ error (95.4\% certainty). 
Projectile breakup enhances $\Gamma_{\rm R}$ from 1.24 to 1.78 
for the surface-type 3BF and from 0.79 to 1.12 for the volume-type 3BF. 
The difference between $\Gamma_{\rm R}=1.78$ and 1.12 comes 
from presence or absence of 
the partial shell inversion. 
Both the partial shell inversion and the projectile breakup effect are thus 
important for the surface-type 3BF. 
The final result $\Gamma_{\rm R}=1.78$ 
with the partial shell inversion and the 
projectile breakup effect is consistent with the experimental value 
$\Gamma_{\rm R}^{\rm exp}=2.0 \pm 0.8$. 

\begin{table}[htbp]
\caption{Summary of odd-even deviation parameters $\Gamma_{\rm R}$. 
}
\begin{tabular}{c|cc||cc}
\hline \hline
Models&$\Gamma_{\rm R}$(DFM)&
$\Gamma_{\rm R}$(CDCC)&$\Gamma_{\rm R}^{\rm exp}$\\
\hline
 III&0.79&1.12&\\
 IV&1.24&1.78&2.0$\pm$0.8\\
\hline \hline
\end{tabular}
\label{Table-summary-staggering}
\end{table}

\section{Summary}
\label{Summary}
we reanalyzed $^{14,15,16}$C scattering from a $^{12}$C target 
at 83~MeV/nucleon with the Melbourne $g$-matrix DFM and studied 
the mechanism underlying the strong odd-even deviation of 
$\Gamma_{\rm R}^{\rm exp}=2.0 \pm 0.8$. 
We introduced a surface-type 3BF 
to the $^{14}$C~+~$n$~+~$n$ three-body model 
in order to describe the suppression of the s-wave spectroscopic factor 
from $\lambda_s(^{15}{\rm C})= 0.97 \pm 0.08$ to
 $\lambda_s(^{16}{\rm C})=0.35 \pm 0.2$. 
The 3BF allows the single-particle energies 
of the $^{14}$C~+~$n$ subsystem to depend 
on the position $r$ of the second neutron from the center of mass 
of the $^{14}$C~+~$n$ subsystem. 
The 1s$_{1/2}$ orbital is lower than the 0d$_{5/2}$ orbital 
for large $r$, but the shell inversion is restored for small $r$. 
The suppression of $\lambda_s(^{16}{\rm C})$ due to 
the partial shell inversion is stronger than that due to pairing correlation. 
Also for ${\bar r}$($^{16}$C), $\Gamma_{\rm rds}$ and $\Gamma_{\rm R}$, 
anti-halo effects 
due to the partial shell inversion are more important 
than the pairing anti-halo effect and 
make the matter radius of $^{16}$C smaller than that of $^{15}$C and 
eventually enhances $\Gamma_{\rm R}$ largely. CDCC calculations 
with effects of projectile breakup and the partial shell inversion yield 
$\Gamma_{\rm R}=1.72$ that is consistent with the experimental value 
$\Gamma_{\rm R}^{\rm exp}=2.0 \pm 0.8$. We therefore conclude that 
both the partial shell inversion and the projectile breakup effect 
in addition to pairing (di-neutron) correlations are important 
to describe the strong odd-even deviation of 
$\Gamma_{\rm R}^{\rm exp}=2.0 \pm 0.8$.

The parameter $\Gamma_{\rm R}^{\rm exp}$ is also large 
for $^{30,31,32}$Ne~+~$^{12}$C 
scattering at 240~MeV/nucleon~\cite{Takechi12}. 
Projectile breakup effects are small 
because of the high incident energy \cite{Minomo:2011bb}. Meanwhile, 
the shell inversion may take place also in $^{31}$Ne
\cite{Minomo:2011bb}. This suggests that 
the partial shell inversion takes place also 
for $^{31}$Ne. If reaction cross sections are measured systematically 
at higher incident energies such as
240~MeV/nucleon, the $\Gamma_{\rm R}$ determined from the measurement
may detect presence or absence of the partial shell
inversion. Systematic measurements of $\Gamma_{\rm R}$ and accurate
theoretical analyses of measured $\Gamma_{\rm R}$ are quite interesting
as a future work.


\noindent
\begin{acknowledgments}
The authors would like to thank Fukuda and Yamaguchi 
for helpful discussions. 
M. Y. is supported by Grant-in-Aid for Scientific Research (No. 26400278) 
from the Japan Society for the Promotion of Science (JSPS). 
\end{acknowledgments}



\end{document}